\documentclass[11pt,a4paper]{article}

\usepackage{jinstpub} 

\usepackage[utf8]{inputenc}  
\usepackage{isotope}
\usepackage{booktabs}
\usepackage{lineno}
\usepackage{sidecap}
\usepackage[separate-uncertainty = true, group-minimum-digits = 4]{siunitx}

\title{Suppression of alpha-induced lateral surface events in the COBRA experiment using CdZnTe detectors with an instrumented guard-ring electrode}

\collaboration{The COBRA collaboration}

\author[a,b]{J.-H.~Arling}
\author[a]{M.~Gerhardt} 
\author[a]{C.~G\"o\ss{}ling} 
\author[c]{D.~Gehre} 
\author[a,1]{R.~Klingenberg\note{Deceased 24 May 2017}} 
\author[a]{K.~Kr\"oninger} 
\author[a]{C.~Nitsch} 
\author[a]{T.~Quante} 
\author[c]{K.~Rohatsch} 
\author[a]{J.~Tebr\"ugge} 
\author[a,2]{R.~Temminghoff \note{Corresponding author}} 
\author[a]{R.~Theinert} 
\author[c]{S.~Zatschler} 
\author[c]{K.~Zuber} 

\affiliation[a]{Technische Universit\"at Dortmund, Lehrstuhl f\"ur Experimentelle Physik IV\\ Otto-Hahn-Str.~4a, 44221~Dortmund}
\affiliation[b]{Deutsches Elektronen-Synchrotron DESY\\ Notkestrasse~85, 22607~Hamburg} 
\affiliation[c]{Technische Universit\"at Dresden, Institut f\"ur Kern- und Teilchenphysik\\ Zellescher Weg~19, 01069~Dresden}

\emailAdd{robert.temminghoff@tu-dortmund.de}

\abstract{
  The COBRA collaboration searches for neutrinoless double beta-decay ($0\nu\beta\beta$-decay) using
  CdZnTe semiconductor detectors with a coplanar-grid readout and a surrounding 
  guard-ring structure.
  The operation of the COBRA demonstrator at the Gran Sasso underground laboratory (LNGS) indicates 
  that alpha-induced lateral surface events are the dominant source of background events.
  By instrumenting the guard-ring electrode it is possible to suppress this type of
  background.
  In laboratory measurements this method achieved a suppression factor of alpha-induced lateral 
  surface events of $5300^{+2660}_{-1380}$, while retaining \SI{85.3\pm0.1}{\percent} of gamma events 
  occurring in the entire detector volume.
  This suppression is superior to the pulse-shape analysis methods used so far in COBRA by three orders of magnitude.}

\keywords{CZT, CdZnTe, semiconductor detector, coplanar-grid, guard ring, background suppression, surface events}

\arxivnumber{1701.07432 }

\begin{document}

\maketitle

\flushbottom 



\section{Introduction}
The COBRA (\textbf{C}dZnTe \textbf{0} Neutrino Double \textbf{B}eta \textbf{R}esearch \textbf{A}pparatus) collaboration \cite{Zuber:2001vm} searches for 
neutrinoless double beta-decay ($0\nu\beta\beta$-decay) \cite{Furry}.
This decay is forbidden in the Standard Model of particle physics.
The quest for lepton-number violation is a main motivation to search for $0\nu\beta\beta$-decay.
Furthermore, the detection of this process could give information about several general properties of neutrinos 
like the neutrino mass scale and mass hierarchy.
The decay has not been measured yet, limits on the half-life are of the order of \SI{E25}{yr}, 
depending on the nuclide under study \cite{Agostini:2015, Asakura:2014, Alduino:2016}.
Special experimental techniques are required to measure such rare decays.
One crucial issue is the reduction of background events which can mimic the searched-for decay, 
coining the term 'low-background experiment'.

The COBRA collaboration operates a demonstrator setup at the Gran Sasso underground 
laboratory (Italy), technical details can be found in Ref.~\cite{COBRA_demonstrator}. 
The demonstrator consists of 64 CdZnTe coplanar-grid (CPG) semiconductor detectors with the 
volume of \SI{1}{\cubic\centi\metre} each. 
CdZnTe contains nine nuclides that can undergo double beta-decays: \isotope[64]{Zn}, 
\isotope[70]{Zn}, \isotope[106]{Cd}, \isotope[108]{Cd}, \isotope[114]{Cd}, \isotope[116]{Cd}, 
\isotope[120]{Te}, \isotope[128]{Te} and \isotope[130]{Te}.
Due to its high $Q$-value of \SI{2814}{\kilo\electronvolt} \cite{Rahaman}, \isotope[116]{Cd} is 
the most promising candidate for the COBRA experiment.
No significant excess over the estimated background was found which results in half-life limits of about 
\SI{E21}{yr} for the ground-state to ground-state $0\nu\beta^-\beta^-$ transitions of
\isotope[70]{Zn}, \isotope[114]{Cd}, \isotope[116]{Cd}, \isotope[128]{Te} and \isotope[130]{Te}
as documented in Ref.~\cite{COBRA_analysis}.
One conclusion drawn from the operation of the demonstrator is that the main background component 
stems from alpha-induced lateral surface events. 
In the current scheme, lateral surface events are identified by analyzing the recorded pulse shapes in the 
detector; details of this method can be found in Ref.~\cite{COBRA_PSA}.

All detectors used by COBRA feature a guard ring, which is a boundary electrode surrounding the CPG anodes. 
It is a common method to improve the detector performance, as it leads 
to a better balanced weighting potential \cite{He1998} and to a reduction of leakage currents.
In the current configuration, the guard ring is not instrumented and left on a floating potential.
Setting a defined potential on a guard-ring structure to suppress surface events was studied in 
Ref.~\cite{EDELWEISS} using Germanium detectors. First measurements have shown that this novel method can
be used for the COBRA detectors as well without deteriorating the detector performance significantly \cite{tebruegge_diss}.
This paper discusses the instrumentation of guard rings for large 
$\SI{6}{\cubic\centi\metre}$ detectors used for the upgrade of the COBRA demonstrator.

\section{Guard-ring instrumentation of CdZnTe detectors}
\label{sec:methods}
\begin{SCfigure}[][!htb]
  \centering
    \includegraphics[width=0.5\textwidth]{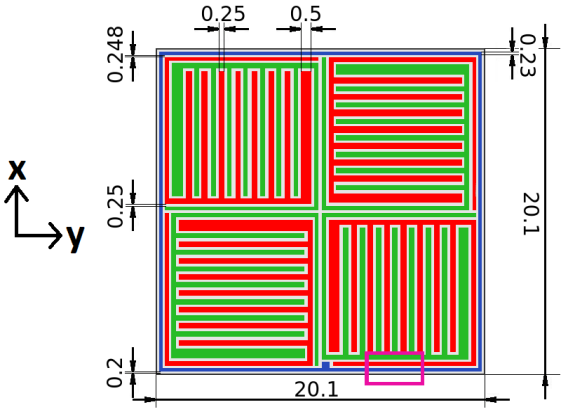}
    \caption{Scheme of the electrode configuration of a CPqG detector. The four sectors comprise individual 
	     CPGs, each instrumenting a collecting anode (CA, red) and a non-collecting anode 
	     (NCA, green). The surrounding guard ring (GR) is 
	     shown in blue. All distances are given in \si{\milli\metre}. The purple box indicates the
	     position of the detailed view shown in \autoref{fig_anode_alpha}.}
    \label{fig_sketch_cpqg}
\end{SCfigure}

In CdZnTe, the product of lifetime and mobility for electrons and holes is different by three orders of 
magnitude, introducing a strong interaction-depth dependence of the detectors response.
To compensate for this effect, the so-called coplanar-grid technology \cite{cpg94} is 
used in COBRA. 
It is a single-polarity charge-carrier sensing method, where in this case only the electron 
signal is read out.
In this technique, two anodes are comb-shaped interleaved and are set on a slightly 
different electric potential, called grid bias (GB).
One anode is set to ground potential and referred to as collecting anode (CA) as it 
collects the generated electrons. The other is set on a negative potential of typically 
\SI{-40}{\volt} and is referred to as non-collecting anode (NCA). 
A bias voltage (BV) of typically \SI{-1200}{\volt} is applied to the cathode.
The detectors discussed in this publication are $(20\times20\times15)\,$\si{\cubic\milli\metre} 
in size with a volume of \SI{6}{\cubic\centi\metre} and a mass of \SI{36}{\gram}.
Their electrodes are configured according to a so-called coplanar quad-grid (CPqG), i.e.~four individual CPG structures,
rotated against each other by \SI{90}{\degree}. The whole CPqG structure is surrounded by a common
guard ring (GR) which is left on a floating potential in a default configuration.
Such a detector, but without a guard-ring electrode, was characterized extensively in 
Ref.~\cite{COBRA_quad}.
Details of the electrode design and its dimensions are depicted in \autoref{fig_sketch_cpqg}. 

For the instrumentation of the guard ring a defined potential is set to it, in this case ground potential, 
which is the same potential as the CA. The guard ring can thus collect charges drifting to it.
This is particularly important for charge clouds originating from interactions in the vicinity of the 
lateral detector surfaces. 
The anodes and the guard ring are connected to charge sensitive preamplifiers (Cremat CR110\footnote{\url{http://www.cremat.com/CR-110.pdf}}). 
These convert the induced charge signals coming from interactions in the 
detector volume into voltage signals.
The signals are afterwards amplified in linear amplifiers and digitized in flash analog-to-digital 
converters (FADC SIS3300\footnote{\url{http://www.struck.de/sis3300.htm}}) with a sampling rate of \SI{100}{MHz}. The data contains the full
pulse-shape information of the measured channels, making it possible to apply pulse-shape analysis
tools in the off-line analysis \cite{COBRA_PSA}. More details about the data-acquisition can be 
found in Ref.~\cite{COBRA_demonstrator}.

The deposited energy in a CPG detector can be reconstructed by calculating the amplitude $A$ of the difference signal 
between the CA and NCA signal amplitudes including a so-called weighting factor $w$,
\begin{equation}
 E \propto A_{\text{CA}} - w \cdot A_{\text{NCA}}.
 \label{eqn:E}
\end{equation}
The weighting factor compensates for effects of electron trapping \cite{cpg94} and can be 
determined during the calibration process.
The energy deposited in each CPqG sector can be reconstructed individually using \autoref{eqn:E}.
The signal of the guard ring is not considered in the energy reconstruction in this study.

\section{Charge-cloud dynamics of alpha particles}
\label{sec:alphas}
The aim of this chapter is to roughly estimate the size of the charge cloud based on theoretical calculations
to show that from this point of view it is worth to investigate the guard-ring instrumentation for vetoing surface events.
The penetration depth of alpha particles from radioactive decays in solids like CdZnTe is very 
short, about $R_\text{initial}=$\SI{20}{\micro\metre} for an alpha particle with a kinetic energy of \SI{5}{\mega\electronvolt}.
Therefore, events induced by sources of alpha radiation outside the detector volume deposit their energy
in the vicinity of the surfaces and are referred to as lateral surface events.
Due to the deposited energy, electron-hole pairs are created at the interaction point. 
The effect of holes is ignored here because their drift towards the cathode is much slower.
The charge cloud of the generated electrons expands while drifting through the detector towards the electrodes 
due to two main reasons: mutual repulsion of the electrons as well as thermal diffusion.
The maximal values for these two quantities can be calculated by considering near-cathode events
in the detector. In this case, the created electron cloud drifts through the whole detector
volume.

The spread $\sigma_\text{diff}$ of the charge cloud due to diffusion \cite{Knoll} can be 
calculated as a function of the drift length $x$ as
\begin{equation}
 \sigma_\text{diff}(x) = \sqrt{\frac{2 k_B T x}{e E}},
\end{equation}
where $k_\text{B}$ is the Boltzmann constant, $T$ the absolute temperature, $e$ the electric 
elementary charge and $E$ the electric field strength.
To estimate the diffusion and to better compare it with the repulsion, an interval of $3\,\sigma$ containing about \SI{99.7}{\%} of all charges is used.
At room temperature and with an applied BV of \SI{-1200}{\volt} the resulting charge-cloud expansion due to
thermal diffusion for a maximal drift length of \SI{15}{\milli\metre} is 
\begin{equation}
 R_\text{diff}^\text{max}=3\sigma_\text{diff}^\text{max} = 3\sigma_\text{diff}(\SI{15}{\milli\metre}) \approx\,\SI{300}{\micro\metre}.
\end{equation}
The expansion $R_\text{rep}$ due to mutual repulsion is defined as the largest diameter of the charge cloud contained in a sphere \cite{Gatti}.
It can be calculated as function of the drift length to
\begin{equation}
 R_\text{rep}(x) = \sqrt[3]{ \frac{3 e N x} {4 \pi \epsilon_0 \epsilon_r E} }, 
\label{eq_repulsion}
\end{equation}
with the number of created charge carriers $N$ and the permittivity of free space and the relative permittivity, 
$\epsilon_0$ and $\epsilon_r=10.9$ \cite{Capper}, respectively. 
As the charges are very close to the detector wall here, they can only expand into a half-sphere.
Hence, they are concentrated in a smaller volume.
The diffusion is independent from any direction, but for the repulsion one can conservatively estimate the effect by accounting only a half-sphere as containing volume. 
Therefore, the repulsion can be corrected by the following replacement in \autoref{eq_repulsion}:
\begin{equation}
R_\text{rep} \propto \sqrt[3]{\frac{N}{4\pi}} \rightarrow R_\text{rep}^{\text{corr}} \propto \sqrt[3]{\frac{N}{2\pi}} \approx 1.25.
\end{equation}
Hence, the value for the repulsion is enlarged by about \SI{25}{\%}.
The energy of the alpha particles from the \isotope[241]Am source is about \SI{5.5}{MeV}, which is also
typical for alpha particles from natural decay chains.
In an interaction of this energy, about $10^6$ charge carriers are produced. 
Therefore, the resulting expansion due to repulsion is at most 
\begin{equation}
R_\text{rep}^\text{corr, max}= R_\text{rep}^\text{corr}(\SI{15}{\milli\metre}) \approx\,\SI{550}{\micro\metre}.  
\end{equation}
The quadratic sum of these two effects and the initial penetration depth can be calculated to estimate an upper limit on the maximal 
spread $L_\text{max}$ of a charge cloud \cite{Donmez, Benoit} assuming a maximal drift length of 
\SI{15}{\milli\metre} for near-cathode events,
\begin{equation}
  L_\text{max} = \sqrt{ \left(R_\text{diff}^\text{max}\right)^2 + \left(R_\text{rep}^\text{corr, max}\right)^2 + \left(R_\text{initial}\right)^2} \approx \SI{620}{\micro\metre}.
 \label{eq_L_max}
\end{equation}

\begin{SCfigure}[][!htb]
  \centering
    \includegraphics[width=0.42\textwidth]{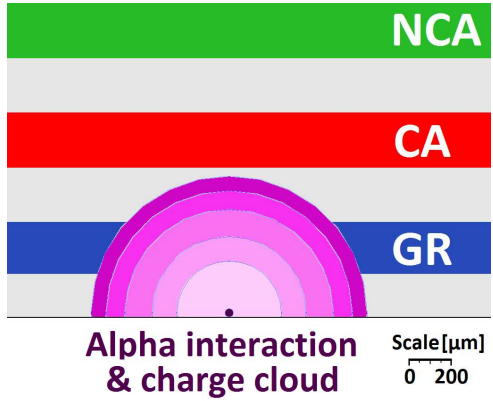}
\caption{Detailed top view of the anode side as marked in
	 \autoref{fig_sketch_cpqg}. The different types of magenta indicate the 
	 expansion of the charge cloud after a drift length of \SIlist{1;3;6;10;15}{\milli\metre}. 
	 The dark dot highlights the initial charge cloud.}
    \label{fig_anode_alpha}
\end{SCfigure}

The effect of the charge-cloud expansion of alpha-induced lateral surface events on the electrodes
is shown in \autoref{fig_anode_alpha}.
The initial interaction has a distance of about $\SI{20}{\micro\metre}$ from the surface. 
The expansion of the charge cloud is shown after a drift length of \SIlist{1;3;6;10;15}{\milli\metre}.
Even for the largest drift length of \SI{15}{\milli\metre}, the outermost CPG anode (CA) is not affected 
by the charge cloud.
Hence, a clear separation of alpha-induced lateral surface events and events occurring in the inner detector 
volume should be possible by instrumenting the guard ring.
This is investigated in the following.

\section{Predictions from an electric-field calculation}
\label{sec_field_simulations}
To study the effect of different guard-ring potentials, e.g.~floating potential in the default setting
or ground potential for enabling charge collection, an electric-field
simulation is performed using the simulation tool COMSOL Multiphysics in version 5.2 \cite{comsol}.

\begin{SCfigure}[][!htb]
  \centering
    \includegraphics[width=0.53\textwidth]{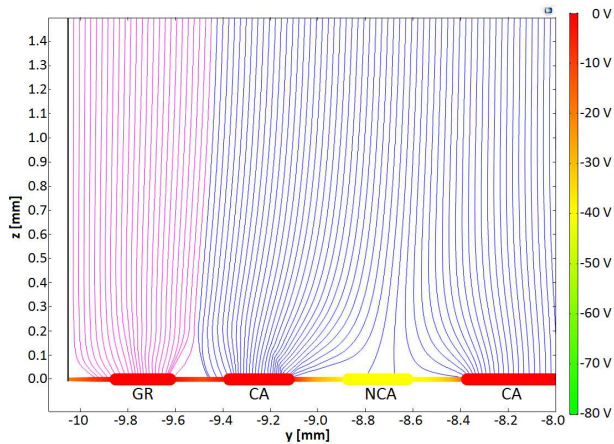}
    \caption{Electric field-line distribution shown in a cross section close to the detector edge. The guard ring is set on 
    ground potential. The chosen GB and BV are \SI{-40}{\volt} and \SI{-1200}{\volt}. Field
    lines ending on the GR are colored in pink, those ending on the CPqG anodes in blue.}
    \label{fig_fieldlines}
\end{SCfigure}

The current CPqG detector design is implemented in terms of dimensions and electrode design,
the typical material-specific values for CdZnTe are taken from Ref.~\cite{Zhang}. 
The applied bias voltage is assumed to be \SI{-1200}{\volt} and a typical value of \SI{-40}{\volt} 
is chosen for the GB. Configurations with different guard-ring potentials are simulated. 
In addition, the configuration of the CPqG anodes is varied: if a CA or an NCA 
bias is applied to the outermost anode, the configuration is referred to as $\text{CA}_{\text{out}}$ and $\text{NCA}_{\text{out}}$ mode, respectively.
The simulated electric field-line distribution for the $\text{CA}_{\text{out}}$ mode with the guard ring 
on ground potential is shown in \autoref{fig_fieldlines}.
Here, field lines start on the cathode in equidistant steps of \SI{25}{\micro\metre} and they end on the 
different electrodes (GR, CA and NCA as highlighted).
Field lines ending on the guard ring are colored in pink in contrast to field lines ending on either of the 
CPqG anodes, which are shown in blue.

\begin{SCfigure}[][!htb]
  \centering
    \includegraphics[width=0.4\textwidth]{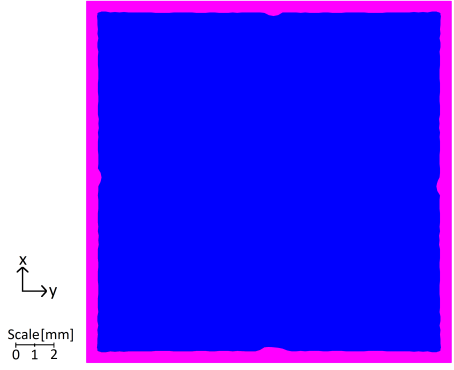}
    \caption{Contour plot of charge collection of the guard ring and the CPqG anodes in $\text{CA}_\text{out}$ 
	     mode as seen from the top.
	     Shown are the projected start positions of field lines finally ending on the 
	     guard ring (pink) or on either of the CPqG anodes (blue). The chosen GB is \SI{-40}{\volt} and BV 
	     is \SI{-1200}{\volt}.}
    \label{fig_contour}
\end{SCfigure}

The innermost field line ending on the guard ring starts at a distance of \SI{625}{\micro\metre} from the 
detector edge. The maximal distance depends on the depth of the starting position, and a slope for the first 
\SI{2}{\milli\metre} below the anode side is observed before the field lines are almost parallel.
Energy deposited between the detector surface and the innermost field line ending on the guard ring will not 
be detected by the CA.
This will lead to a decrease in the overall detection efficiency, but it will in particular suppress
a large fraction of alpha-induced lateral surface events, according to \autoref{eq_L_max}.
The area influenced by the instrumented guard ring is visualized in \autoref{fig_contour}. 
This contour plot depicts the start positions of electric field lines and differentiates between 
end positions on the guard-ring electrode or the CPqG anodes. 
The dips in the area influenced by the CPqG at the middle of the sides arise from the transitions between the 
different CPG sectors. At the bottom, the dip is larger because of the guard-ring contact pad.
The ratio of the area influenced by the guard ring relative to the full area is a measure 
of reduction of the fiducial detector volume, $\epsilon_\text{fid}$. 
The results for the $\text{CA}_{\text{out}}$ mode, as depicted in \autoref{fig_contour}, and for the 
$\text{NCA}_{\text{out}}$ mode are 
\begin{equation}
 \epsilon_\text{fid}^{\text{CA}_{\text{out}}}=\SI{87.7}{\percent} \qquad\text{and}\qquad \epsilon_\text{fid}^{\text{NCA}_{\text{out}}}=\SI{86.0}{\percent},
\label{eq_eta_simulation}
\end{equation}
respectively. The uncertainties in these calculations are negligible as the mesh for the numerical calculation 
is fine (\SI{25}{\micro\metre}), and small variations of the biases do also not show a notable effect.\\
Only the $\text{CA}_{\text{out}}$ mode is evaluated in the following because of the expected larger efficiency than in $\text{NCA}_{\text{out}}$ mode.
In addition, other beneficial effects like less charge-sharing between the sectors favor this mode \cite{Temminghoff}.

\section{Experimental setup}
\label{sec:measurement}
The effect of instrumenting the guard ring in terms of a suppression of alpha interactions from
the lateral surfaces is studied in measurements.
A prototype CPqG detector with guard ring, produced by Redlen Technologies\footnote{\url{http://redlen.ca/}}, is used. 
During characterization of this test detector, the energy resolution in terms of full-width at half maximum (FWHM) for an incident energy of \SI{662}{\kilo\electronvolt} of the different CPG sectors is measured. 
Two of the four sectors have only poor energy resolutions of \SI{5.0}{\percent} and 
\SI{7.0}{\percent}.
Another sector is suffering from leakage currents degrading its performance immensely, while not 
influencing the other sectors.
Finally, the last remaining sector provides a satisfactory energy resolution of 
\SI{3.0}{\percent}. This sector is therefore chosen as the sector for testing.

A \isotope[241]{Am} source emitting alpha particles with an energy of 
\SI{5.5}{\mega\electronvolt} is used to provide a sample of alpha-induced surface interactions. 
The source is pointed directly on the sector under test. 
Furthermore, the detector can be irradiated with gamma radiation from a \isotope[232]{Th} 
source with various gamma lines, the highest at an energy of \SI{2614}{\kilo\electronvolt}. 
As the low-energy threshold of the measurement is at about \SI{280}{\kilo\electronvolt} and the detector used here has a side-length of \SI{2}{\centi\metre}, photon interactions can take place all throughout its bulk, as even the lowest energy photons considered here have a considerable probability to pass through the detector.

\begin{figure}[!htb]
  \centering
    \includegraphics[width=0.9\textwidth]{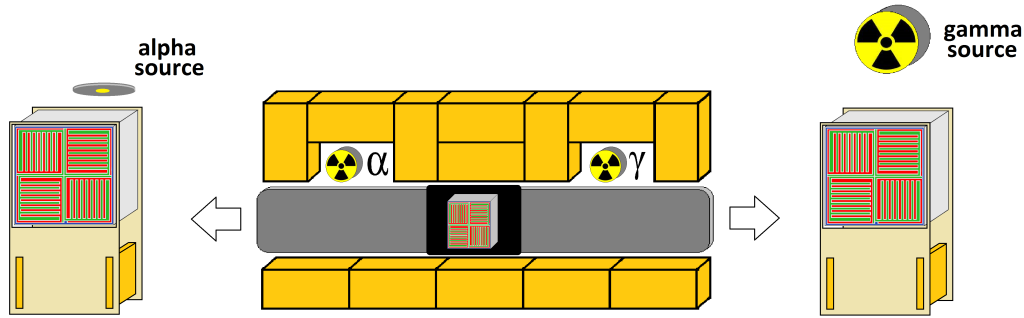}
    \caption{Top view of the measurement setup for evaluating the effect of the guard-ring instrumentation.
    The detector is mounted on a movable stage allowing to shift it between the two different
    radioactive sources ($\alpha$: \isotope[241]{Am} source, $\gamma$: \isotope[232]{Th} 
    source). The alpha source is pointed to the sector under test in a distance of approximately \SI{7}{\milli\meter}, whereas the 
    gamma source irradiates the whole detector. The yellow bricks symbolize lead shielding.}
    \label{fig_setup}
\end{figure}

The detector is mounted on a movable stage, so that it is possible to irradiate the detector 
at first with gamma radiation and afterwards with the alpha source without switching off the biases
when exchanging the sources.
Hence, the recorded gamma energy spectrum can be used for a common calibration for each 
configuration. 
The measurement setup as well as the tested sectors are depicted in \autoref{fig_setup}.

To account for background, a dedicated measurement of the background spectrum is performed.
The detector is kept in the same position as for an alpha measurement, but the 
\isotope[241]{Am} source is removed. 
This background consists mainly of photons emitted by the decay of naturally occurring radioactive isotopes in the laboratory, 
like \isotope[40]{K}, \isotope[232]{Th}, or \isotope[238]{U}, cosmic muons and photons scattered from the \isotope[232]{Th}-source, 
which was present during this measurement to mimic the situation of the alpha measurement as close as possible. Especially at higher 
energies\footnote{This means basically above the \SI{2614}{\kilo\electronvolt.} line of \isotope[208]{Tl}}, muons will be the dominating 
contribution, as it is known that the activity of internal alpha contaminations is very small (less than 0.5 ppb for U and Th). This has been 
measured for the bulk material of the detector by Redlen Technologies. Contaminations by $\alpha$ particles on the surface will lose 
most of their energy already in the coating of the detectors and will thus not contribute to the high energy part, while it was measured that 
the coating itself is clean enough to not contribute to the background significantly at this level. As the aim of COBRA is to have a 
much lower background for the upgrade of the demonstrator setup at the LNGS as it is possible in our laboratory at the surface level, 
this discussion is only valid for the comparatively high-rate measurements discussed in this publication.

To evaluate the effect of the instrumented guard ring, one measurement is performed with
the guard ring on ground potential and one with the guard ring kept on a floating potential.
The applied voltages for this measurement campaign are a GB of \SI{-40}{\volt} and 
a BV of \SI{-1200}{\volt}.

\section{Measurements and results}
\label{sec:results}
\begin{SCfigure}[][!htb]
  \centering
    \includegraphics[width=0.79\textwidth]{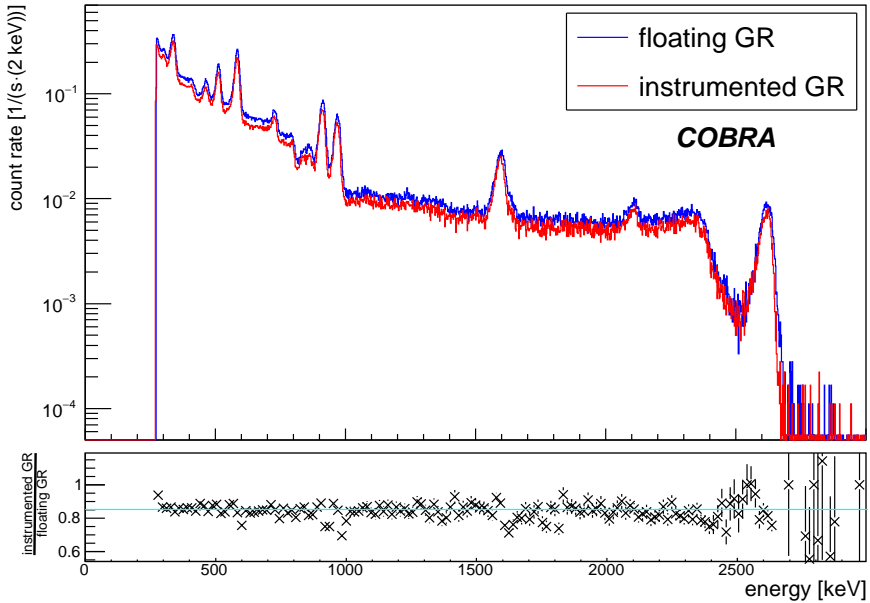}
    \caption{Top: \isotope[232]{Th} energy spectra measured with the sector under test. Shown are 
    the cases of floating (blue) and instrumented guard ring (red). Bottom: Ratio-plot of the two spectra above. The cyan colored constant is the 
    value of $\epsilon_\gamma$ calculated in \autoref{eq_epsilon_gamma}.}
    \label{fig_th232}
\end{SCfigure}

The energy spectra of the sector under test irradiated with \isotope[232]{Th} gamma radiation are shown in \autoref{fig_th232}. 
The reconstructed energy spectrum for the case of a floating guard ring is depicted in blue and 
for the case of an instrumented guard ring in red.
The sharp drop around \SI{280}{keV} stems from the threshold of the energy reconstruction needed for event-by-event triggering.
The spectrum shows the typical features of a \isotope[232]{Th} source.
The total event rate in the case of an instrumented guard ring is reduced 
compared to the floating case due to the reduced efficiency.
This is expected from the results of the electric-field calculation in \autoref{sec_field_simulations}.
The efficiency shows a slight energy dependence.
However, the ratio of the peak contents changes by less than \SI{10}{\%} from \SI{338}{keV} to \SI{2614}{keV}. This is also expected, as the 
lowest energetic photons considered here have a mean free path of about \SI{10}{mm} and hence, only about $\SI{5}{\%}$ of all interactions 
will take place in the volume influenced by the guard ring.
Consequently, more than $\SI{95}{\%}$ of this radiation interacts in the detector center measured by the CPqG. For higher energetic gamma 
radiation the fraction of events interacting near the guard ring is even smaller, but as the probability to interact in the bulk of the 
detector is also reduced, nearly no net effect can be seen in the ratio plot of the two spectra in \autoref{fig_th232}.
The constant value of the reduction of the fiducial volume $\epsilon_\gamma=\SI{85.3\pm0.1}{\percent}$ --- which will be calculated in 
\autoref{eq_epsilon_gamma} --- is indicated in the plot as a cyan line as well.
This shows that while in principle an energy dependence exists, it is so small that it can be neglected here.

\begin{SCfigure}[][!htb]
  \centering
    \includegraphics[width=0.7\textwidth]{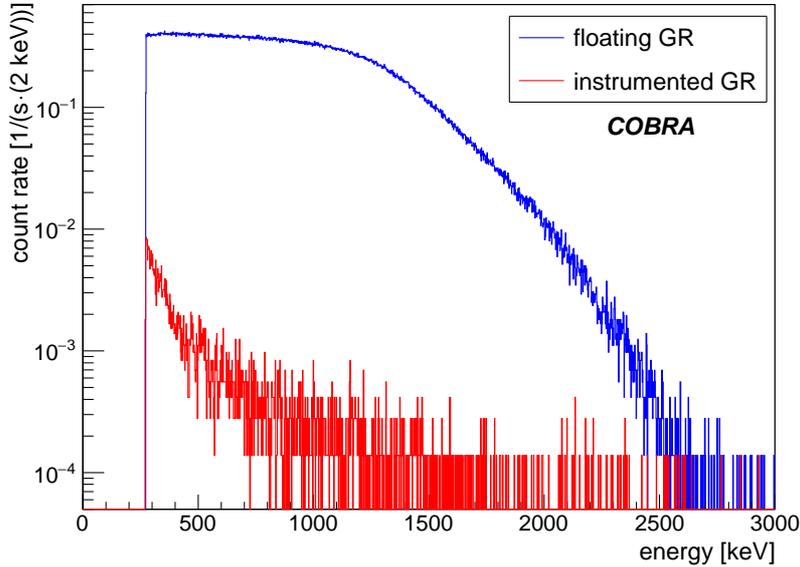}
    \caption{\isotope[241]{Am} energy spectra measured with the sector under test.
            Shown are the cases of floating (blue) and instrumented guard ring (red).}
    \label{fig_am241}
\end{SCfigure}

For the alpha measurement, the resulting energy spectra for both guard-ring configurations are 
shown in \autoref{fig_am241}.
The unusual spectral shape of the alpha spectrum for the floating guard ring can be 
explained as follows:
The source irradiates the detector uncollimated, otherwise small localized crystal effects like inhomogeneities could affect the results.
As the distance between source and detector is only about \SI{7}{\milli\meter}, 
the path the alpha-particles have to travel until they reach the detector surface is different by up a factor of two due to geometric reasons.
Furthermore, the used detectors here are known to have a small dead layer and are coated with an epoxy based resin, which both are not necessarily distributed homogeneously.
The same geometric argument as above is also true for the distance the alpha particles have to travel through the lacquer and the dead layer.
This results in a smeared distribution shifted towards lower energies.
This effect has been verified qualitatively by a Monte-Carlo simulation using GEANT4. For this, a simplified geometry was used and effects by event 
pile-up, background radiation and energy resolution were neglected, but the resulting spectral shape showed the same principal features.
The interesting point here is that the spectrum of alpha-induced lateral surface events in the instrumented guard-ring configuration
is clearly reduced compared to the floating potential case.

\begin{SCfigure}[][!htb]
  \centering
    \includegraphics[width=0.7\textwidth]{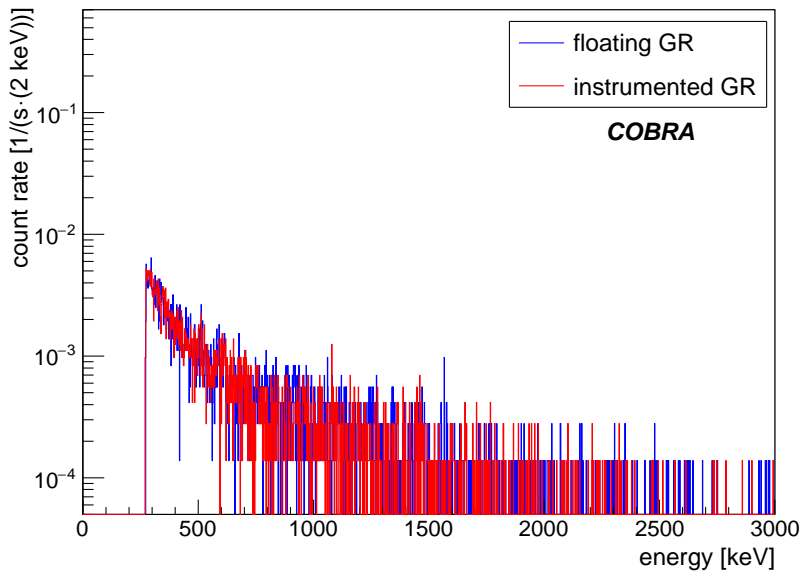}
    \caption{Background energy spectra measured with the sector under test. 
	     Shown are the cases of floating (blue) and instrumented guard ring (red).}
    \label{fig_background}
\end{SCfigure}

The spectra resulting from dedicated measurements of the laboratory background are shown in 
\autoref{fig_background}.
One can see similar spectra for both configurations of the guard ring again. 
This is expected because most background events stem from gamma radiation and muons.
A large fraction of the events in the spectrum from the alpha measurement with an instrumented 
guard ring (red curve in \autoref{fig_am241}) are caused by laboratory background, which needs 
to be considered when estimating the suppression of alpha-induced events.
\newline
This is evaluated quantitatively in \autoref{tab:results} for all six measurements by comparing
the count rates, integrated between \SI{280}{\kilo\electronvolt} and \SI{3000}{\kilo\electronvolt}. 
The quoted uncertainties are due to the statistical Poisson error.
\begin{table}[!h]
\caption{Results of measurements for sector under test. Given are the integral count rates $r$
(between trigger threshold and chosen high energy cut-off) in each measurement and for both guard-ring configurations.}
\label{tab:results}
\vspace{-0.5cm}
\begin{center}
\begin{tabular}{l l c c c }
\toprule
  & & \text{alpha measurement} & \text{background measurement} & \text{gamma measurement} \\
\midrule
$\begin{aligned}& r_{\text{floating GR}} \\& r_{\text{instrumented GR}} \end{aligned}$ &
$\begin{aligned}& \text{[1/s]} \\& \text{[1/s]} \end{aligned}$ &
$\begin{aligned} \num{208.6}&\pm\num{0.2} \\ \num{0.498}&\pm\num{0.008} \end{aligned}$ &
$\begin{aligned} \num{0.493}&\pm\num{0.008} \\ \num{0.465}&\pm\num{0.008} \end{aligned}$ &
$\begin{aligned} \num{42.33}&\pm\num{0.05} \\ \num{36.15}&\pm\num{0.04} \end{aligned}$ \\
\bottomrule
\end{tabular}
\end{center}
\vspace{-0.35cm}
\end{table}

The resulting reduction of the fiducial volume for gamma radiation in the case of an 
instrumented guard ring is estimated as
\begin{equation}
\epsilon_\gamma=\frac{r^{\gamma}_{\text{instrumented GR}}-r^{\text{bkg}}_{\text{instrumented GR}}}{r^{\gamma}_{\text{floating GR}}-r^{\text{bkg}}_{\text{floating GR}}}.
\label{eqn:gamma} 
\end{equation} 
Using the numbers in \autoref{tab:results}, this yields
\begin{equation}
\epsilon_\gamma=\SI{85.3\pm0.1}{\percent}.\\
\label{eq_epsilon_gamma}
\end{equation}
This value is comparable with the prediction of $\epsilon_\text{fid}=\SI{87.7}{\percent}$ from the calculation of 
the electric field (\autoref{eq_eta_simulation}). The relative difference between the calculated and the measured value is 2.7\%, which is larger than 
the difference expected due to purely statistical reasons.
Instead, the difference arises from effects of a real detector that are not incorporated in the electric-field calculation, e.g.~inhomogeneities in 
the detector bulk, uncertainties in the electrode placements or the energy resolution. The simulation did also not take into account the effect of 
multiple scattering of high energetic photons in the detector, which in principle leads to a slightly higher chance for an interaction to take 
place in the volume affected by the guard ring.

The suppression factor for alpha-induced lateral surface events $\text{SF}_{\alpha}$ can be defined as
\begin{equation}
\text{SF}_\alpha=\frac{r^{\alpha}_{\text{floating GR}}-r^{\text{bkg}}_{\text{floating GR}}}{r^{\alpha}_{\text{instrumented GR}}-r^{\text{bkg}}_{\text{instrumented GR}}}.
\label{eqn:alpha} 
\end{equation}
As the rates in the enumerator and the denominator differ by about four orders of magnitude, 
common Gaussian uncertainty propagation cannot be used for the calculation of the ratio.
Instead, the suppression factor is estimated based on a toy Monte-Carlo study: 
\num{E7} pairs of random numbers are drawn from two Gaussian distributions, characterized by
the enumerator and the denominator, and the resulting frequency distribution of the ratios of 
these numbers is interpreted as the probability density of $\text{SF}_{\alpha}$.
Because this asymmetric distribution features a strong non-Gaussian tail towards larger values,
the suppression factor is estimated by the mode of the distribution.
The smallest interval containing \SI{68.3}{\%} of all entries is interpreted as the uncertainty interval. 
This yields
\begin{equation}
\text{SF}_\alpha=5300^{+2660}_{-1380}.
\label{eq_SF_result}
\end{equation}
This is a large improvement compared to the pulse-shape analysis currently used to
suppress lateral surface events in COBRA \cite{COBRA_PSA, COBRA_analysis}: 
the relative detection efficiency for gamma radiation is about the same, but the 
suppression factor for alpha-induced lateral surface events is 
nearly three orders of magnitude larger.

The main region of interest for COBRA is around the $Q$-value of \isotope[116]Cd at \SI{2614}{keV}.
The suppression factor for alpha-induced lateral surface events is calculated here only up to about \SI{2.5}{MeV} for the reasons discussed above.
However, no deterioration of this factor is expected for higher energies.
The charge cloud of a typical alpha particle with an energy of \SI{5}{MeV} should not reach the inner detector volume.
The expansion of the charge cloud of a hypothetic alpha particle with an energy of \SI{10}{MeV}, which is a little more than the prominently occurring natural 
energies of alpha radiation can be estimated analog to \autoref{sec:alphas} to be maximally \SI{750}{\micro\meter} for an interaction close to the cathode.
This charge cloud could reach into the CPqG volume in the detector center partly.
But even this is no drawback to this method, as this type of events could be vetoed in a coincidence analysis easily, as energy is deposited in the guard ring and the CPqG-area at the same time.

\section{Summary and outlook}
\label{sec:conclusion}
The COBRA collaboration searches for neutrinoless double beta-decay and hence background reduction is
a crucial issue.
Alpha-induced lateral surface events are the main background source when operating the COBRA demonstrator.
In this paper, dedicated laboratory measurements and calculations of electric fields 
are used to demonstrate that instrumenting the 
guard ring of CPqG~detectors leads to a suppression factor for alpha-induced lateral surface
events of $5300^{+2660}_{-1380}$, while the reduction of fiducial volume for gamma 
events occurring throughout the entire detector volume is \SI{85.3\pm0.1}{\percent}.\\
This concept will be followed in the upgrade of the COBRA extended demonstrator (XDEM).
Under optimal conditions and the assumption that the background is dominated by alpha events, 
the overall background rate is expected to be lowered by two orders of magnitude.\\
The signals of the guard ring are measured, but not included in the analysis presented here.
A modified event reconstruction including the guard-ring signal can in principle be performed. 
Preliminary studies indicate that using this additional information is feasible and can improve 
the detector performance (e.g.~the detection efficiency) and the veto capabilities for lateral surface events even further.

\acknowledgments 
We thank the LNGS for the continuous support of the COBRA experiment. COBRA is supported by 
the German Research Foundation DFG (ZU 123/15-1 / GO 1133/3-1). Furthermore, we thank COMSOL for support with the
COMSOL Multiphysics simulation and O. Schulz for technical support with the DAQ.

\bibliography{guard_ring_instrumentation.bib}
\bibliographystyle{unsrt}

\end{document}